\documentclass[aps,twocolumn,showpacs,superscriptaddress,amsmath,amssymb,amsfonts,floatfix,longbibliography]{revtex4-1}
\usepackage[T1]{fontenc} 
\usepackage{graphicx}
\usepackage{float}
\usepackage{dcolumn}
\usepackage{bm}
\usepackage{amssymb}
\usepackage{microtype}
\usepackage{xfrac}
\usepackage{gensymb}
\usepackage[most]{tcolorbox}
\usepackage{xcolor}
\usepackage{multirow}
\usepackage{enumitem}
\usepackage{natbib,hyperref}
\usepackage{graphicx}

\hypersetup{
	colorlinks=true, 
	citecolor=blue,  
}

\begin{document}


\title{Phase stability of entropy stabilized oxides with the $\alpha$-PbO$_2$ structure}

\author{Solveig S. Aamlid}
    \affiliation{Stewart Blusson Quantum Matter Institute, University of British Columbia, Vancouver, BC V6T 1Z4, Canada}

\author{Graham H.J. Johnstone}
    \affiliation{Stewart Blusson Quantum Matter Institute, University of British Columbia, Vancouver, BC V6T 1Z4, Canada}
    \affiliation{Department of Physics \& Astronomy, University of British Columbia, Vancouver, BC V6T 1Z1, Canada}

\author{Sam Mugiraneza}
    \affiliation{Stewart Blusson Quantum Matter Institute, University of British Columbia, Vancouver, BC V6T 1Z4, Canada}
    \affiliation{Department of Chemistry, University of British Columbia, Vancouver, BC V6T 1Z1, Canada}

\author{Mohamed Oudah}
    \affiliation{Stewart Blusson Quantum Matter Institute, University of British Columbia, Vancouver, BC V6T 1Z4, Canada}

\author{J\"org Rottler}
    \affiliation{Stewart Blusson Quantum Matter Institute, University of British Columbia, Vancouver, BC V6T 1Z4, Canada}
    \affiliation{Department of Physics \& Astronomy, University of British Columbia, Vancouver, BC V6T 1Z1, Canada}

\author{Alannah M. Hallas}
    \affiliation{Stewart Blusson Quantum Matter Institute, University of British Columbia, Vancouver, BC V6T 1Z4, Canada}
    \affiliation{Department of Physics \& Astronomy, University of British Columbia, Vancouver, BC V6T 1Z1, Canada}
    \email{alannah.hallas@ubc.ca}

\date{\today}
\begin{abstract}

The prediction of new high entropy oxides (HEOs) remains a profound challenge due to their inherent chemical complexity. In this work, we combine experimental and computational methods to search for new HEOs in the tetravalent $A$O$_2$ family, using exclusively $d^0$ and $d^{10}$ cations, and to explain the observed phase stability of the $\alpha$-PbO$_2$ structure, as found for the medium entropy oxide (Ti, Zr, Hf, Sn)O$_2$. Using a pairwise approach to approximate the mixing enthalpy, we confirm that $\alpha$-PbO$_2$ is the expected lowest energy structure for this material above other candidates including rutile, baddeleyite, and fluorite structures. We also show that no other five-component compound composed of the tetravalent cations considered here is expected to form under solid state synthesis conditions, which we verify experimentally. Ultimately, we conclude that the flexible geometry of the $\alpha$-PbO$_2$ structure can be used to understand its stability among tetravalent HEOs.
\end{abstract}

\maketitle

\section*{\label{sec:Introduction}Introduction}
Since the report of the first high entropy oxide (HEO) in 2015, the synthesis of new HEOs has become a topic of extensive research effort, fueled by their many promising potential applications \cite{Rost2015,zhang2019review,musico2020emergent,oses2020high,sarkar2020high,mccormack2021thermodynamics,toher2022high}. While by no means the only definition~\cite{brahlek2022name,doi:10.1021/jacs.2c11608}, a common criterion used to define HEOs is to require that five or more cations should occupy a single lattice site in roughly equimolar proportions. Meanwhile, if only three or four cations share the lattice site it can be considered a medium entropy oxide~\cite{Sarkar2019}. This means that the  phase space for HEO discovery is vast; if elements that cannot reasonably be expected to act as cations in an oxide are excluded, there are 75 cation candidates in the periodic table, which results in more than 17 million possible combinations of five. Although most of these combinations will not form a single phase material, there is no reason to believe that all options are exhausted. There is no tractable experimental approach that can probe this enormous chemical parameter space, yet experimental discovery of new high entropy materials has been the norm. Heuristics, such as the Hume-Rothery and Pauling's rules, allow researchers to estimate if it is reasonable to expect a solid solution to form, based on similarity of crystal structures, ionic radii, electronegativity, and oxidation states of the constituent elements. Accelerating the process of discovery will help decrease the time to application, preferably while targeting some specific functional property. In this endeavor, the combination of chemical intuition, theoretical calculations, and synthesis could help limit the parameter space and identify the most promising compounds before more time and resource intensive exploratory synthesis is attempted. 

Reliably predicting the most stable crystal structure of a compound of any given chemical composition is one of the fundamental efforts in computational materials science, dubbed the Maddox curse~\cite{maddox1988crystals,toher2019unavoidable}. Experimental determination of optimal reaction conditions will always be requisite and the synthesis of metastable phases is certainly possible. Nevertheless, the benefit of achieving predictive power of whether a certain combination of elements is likely to form a stable compound, which crystal structure this compound will prefer, and what the functional properties of the compound would be \textit{a priori} is undoubtedly immense. Over its 60-year history, density functional theory (DFT) has become the gold standard for calculating formation and reaction enthalpies~\cite{hohenberg1964inhomogeneous,kohn1965self}. Considering the successes of high throughput DFT calculations as applied in the Materials Project~\cite{Jain2013} and AFLOW~\cite{curtarolo2012aflow}, databases containing thousands of compounds, one might infer that the curse is lifted. There are success stories, such as in the field of solid electrolytes \cite{Sendek2017}, where screening the database and following up with a funnel approach have lead to the identification of superior materials with certain target properties.  

There are some inherent challenges of DFT applied to HEOs, however, that suggest this approach may not be a panacea. In particular, DFT approaches to formation enthalpies have a few well-known shortcomings, such as representing the structure at zero temperature, overbinding of the oxygen molecule, and failure to predict the correct ground state if there are multiple polymorphs with similar enthalpies~\cite{Wang2006, Curnan2015}. With an increasing number of chemical elements present, predictions from DFT become increasingly less reliable while phase separation or solid solutions become the thermodynamic norm \cite{toher2019unavoidable}. Most entries in databases such as the Materials Project are cation ordered compounds, meaning that each element has its particular symmetrically distinct sublattice(s). Meanwhile, many of the superior functional materials used in industrial applications today are doped or solid solutions, highlighting the necessity of good descriptions of disorder. DFT scales cubically with the number of electrons in the calculation and consequently, since a larger unit cell is needed to accurately represent disorder and the chemical parameter space is so large, a database approach becomes computationally inaccessible for multicomponent solid solutions. This explains why the discovery of new HEOs currently relies on experimental trial-and-error approaches. 

In this work, we explore the phase stability of tetravalent high entropy oxides composed of $d^0$ and $d^{10}$ cations. We repeat the synthesis of a four-component, medium entropy oxide (Ti, Zr, Hf, Sn)O$_2$, the first entropy stabilized oxide with an orthorhombic $\alpha$-PbO$_2$ structure \cite{He2021fourcomponent}. Using DFT, we develop an intuition for why the $\alpha$-PbO$_2$ structure is preferred over the competing rutile, baddeleyite, and fluorite phases. Our analysis reveals that the low symmetry $\alpha$-PbO$_2$ structure is favored due to its structural flexibility, which allows it to accommodate cations of significantly different size. Building on a method developed by Pitike \textit{et al.}~\cite{Pitike2020}, we extrapolate pairwise interaction parameters and evaluate the phase stability of four- and five-component tetravalent oxides. Our calculations suggest that, within the composition and synthesis method space we explore, there are no stable five-component compounds, which we validate experimentally. 

\section*{Results and Discussion}

\begin{figure}
  \includegraphics[width=\columnwidth]{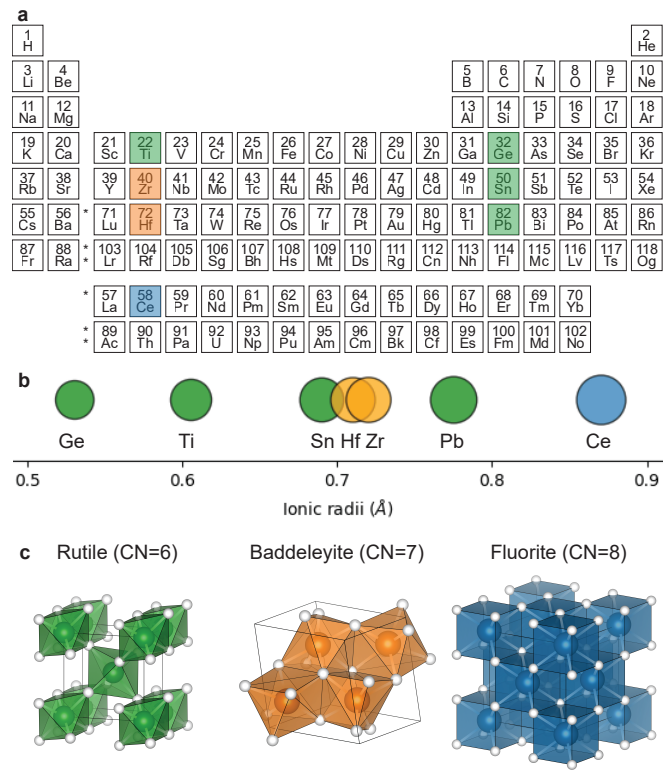}
  \caption{\textbf{Candidate binary oxides.} (a) Periodic table highlighting the tetravalent (4+) cations selected with $d^0$ or $d^{10}$ electronic configurations, color coded according to their ground state crystal structure. The cations under consideration here span three rows of the periodic table and vary dramatically in molar weight. (b) The selected cations organized according to their ionic radii, to scale. (c) The binary oxides $A$O$_2$ of the selected cations have differing ground state crystal structures and coordination numbers (CN). Rutile (green) is found for Ge, Ti, Sn, and Pb, while baddeleyite (orange) is found for Hf and Zf, and fluorite (blue) is found only for Ce.}
  \label{fgr:Itroduction}
\end{figure}

\subsection*{Design intuition for a tetravalent HEO}

Motivated by the search for new entropy stabilized oxides, we chose to explore systems with a single cation sublattice based on tetravalent cations, which until now have not been widely explored. To further narrow our scope, we chose to limit our cations to those that have either $d^0$ or $d^{10}$ electronic configurations, avoiding complications from electronic and magnetic degrees of freedom in the calculations, making them ideal model systems for understanding the mechanism of their entropy stabilization. These criteria lead to the selection of titanium (Ti), germanium (Ge), zirconium (Zr), tin (Sn), cerium (Ce), hafnium (Hf), tin (Sn), and lead (Pb) as potentially suitable tetravalent cations, as marked in the periodic table in Figure \ref{fgr:Itroduction}(a). The elements are distributed across the $p$, $d$, and $f$ blocks and belong to three different rows in the periodic table, resulting in a large variance in their molar masses and ionic radii. The molar masses, which are spanned by the lightest cation Ti ($M=47.9$~g/mol) to the heaviest cation Pb ($M=207.2$~g/mol), vary by a factor of four. In order to make a direct comparison of the ionic radii for these tetravalent cations, we consider their Shannon ionic radius in an octahedral local environment~\cite{Shannon}, which spans from the smallest cation Ge ($r=0.53$~\AA) to the largest Ce ($r=0.87$~\AA) as shown in Figure~\ref{fgr:Itroduction}(b). Silicon (Si) was excluded here because it tends to form ordered compounds when mixed with the other elements due to its smaller size and tendency for tetrahedral oxygen coordination, which as will be discussed later is unfavorable for the formation of a high entropy phase. All compositions considered in this work are equimolar, and we use notations such as (A1, A2, ..., AN)O$_2$ in the chemical formulas to indicate equimolarity.

The most stable crystal structure for the binary oxides of the seven selected elements are rutile (TiO$_2$, GeO$_2$, SnO$_2$, and PbO$_2$), baddeleyite (ZrO$_2$ and HfO$_2$); and fluorite (CeO$_2$), with the the unit cells of each shown in Figure~\ref{fgr:Itroduction}(c). The fluorite structure has an eight-fold coordinated cubic oxygen environment and these cubes are arranged into an edge-sharing motif resulting in a high symmetry cubic crystal structure (space group $Fm\bar{3}m$, No. 225). The baddeleyite structure, meanwhile, is a low-symmetry monoclinic sub-structure of fluorite with an unusual seven-coordinated
oxygen coordination (space group $P2_1/c$, No. 14). Finally, the tetragonal rutile structure has a six-fold coordinate oxygen environment and the structure is composed of corner- and edge-sharing octahedra (space group $P4_2/mnm$, No. 136).

\begin{figure*}
  \includegraphics[width=\textwidth]{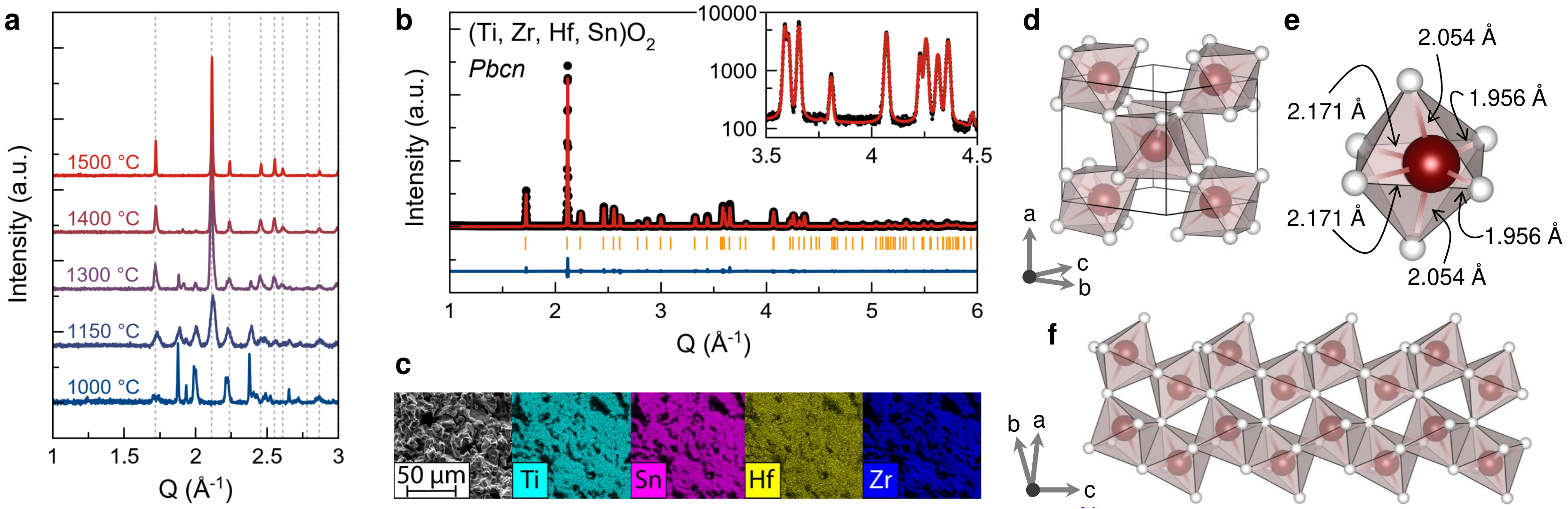}
  \caption{\textbf{Synthesis, crystal structure, and elemental homogeneity of (Ti, Zr, Hf, Sn)O$_2$.} (a) Formation of the single phase (Ti, Zr, Hf, Sn)O$_2$ with the $\alpha$-PbO$_2$ structure from the constituent elemental oxides occurs at a synthesis temperature of 1500 $^\circ$C. Dashed vertical lines indicate the allowed Bragg peak position for this structure. (b) Rietveld refinement (red line) of the powder x-ray diffraction data for (Ti, Zr, Hf, Sn)O$_2$ (black circles) showing excellent agreement with the $\alpha$-PbO$_2$ structure (space group $Pbcn$, No. 60) and no impurity peaks. The residual is given by the blue line and the Bragg peak positions are given by the yellow vertical lines. (c) Scanning electron micrograph and energy dispersive x-ray spectroscopy maps for (Ti, Zr, Hf, Sn)O$_2$ confirming elemental homogeneity at the micron length scale. Each color represent one element. (d) The orthorhombic unit cell for the refined $\alpha$-PbO$_2$ crystal structure. The white atoms are oxygen and the red atoms are the cations. (e) The six-fold coordinate distorted octahedral oxygen environment is highly anisotropic with three inequivalent metal-oxygen bond distances. (f) The structure is composed by a network of corner-and edge-sharing octahedra, with each oxygen shared between three polyhedra.}
  \label{fgr:xray}
\end{figure*}

There are several features of this $A$O$_2$ family that are favourable indicators for the possibility of forming an entropy stabilized phase. The first of these is their extensive polymorphism, suggestive of an inherent structural flexibility. This is most strikingly demonstrated by TiO$_2$, which, in addition to its thermodynamically stable rutile ground state, has eight additional known metastable polymorphs, an indicator that many different structures lie close together in energy~\cite{zhu2014stability}. Likewise, GeO$_2$, HfO$_2$, and PbO$_2$ each naturally occur in one or more metastable phases in addition to their thermodynamic ground states~\cite{micoulaut2006structure,pathak2020structural,white1961high}, and both ZrO$_2$ and SnO$_2$ can be tuned between multiple polymorphic structures by varying temperature and pressure~\cite{endo1990x,smith1965crystal}. Among the ions considered here, only CeO$_2$ is not known to exhibit extensive polymorphism. A second favorable aspect of the $A$O$_2$ family for the successful synthesis of a high entropy compound is the absence of cation ordered phases in the ternary phase diagrams between the binary oxides ($A$O$_2$-$B$O$_2$). Investigation of the 21 constituent ternary phase diagrams in the Materials Project database~\cite{Jain2013} reveals only two such ordered compounds, zircon-like structured ZrGeO$_4$ and HfGeO$_4$ with tetrahedrally coordinated Ge. There are also stable compounds where one cation, typically Pb, has undergone a change in oxidation state, such as pyrochlore Ce$_2$Zr$_2$O$_7$; perovskite PbTiO$_3$, PbZrO$_3$, PbHfO$_3$; and with various symmetries Ti$_3$PbO$_7$, Ge$_3$PbO$_7$, Ge$_3$Pb$_5$O$_{11}$, GePbO$_3$, and Sn(PbO$_2$)$_2$.
Although not completely absent, the small number of cation ordered phases makes this materials system a promising avenue for forming a high entropy solid solution.

\subsection*{Synthesis of a four-component entropy stabilized oxide}
Based on these design criteria, the synthesis of a number of multicomponent high entropy oxides was attempted from various combinations of the tetravalent oxides, leading to the independent discovery of the medium entropy oxide (Ti, Zr, Hf, Sn)O$_2$, previously reported by He \textit{et al.}~\cite{He2021fourcomponent}. This material was prepared by solid state reaction and begins to form at reaction temperatures as low as 1150~$^\circ$C and reaches a single phase after reacting at 1500~$^\circ$C, as shown in Fig.~\ref{fgr:xray}(a). We were also able to achieve a single phase material through repeated firings at  1400~$^\circ$C with intermediate regrinding. (Ti, Zr, Hf, Sn)O$_2$ crystallizes in the orthorhombic $\alpha$-PbO$_2$ structure (space group $Pbcn$, No. 60), a structure that takes its name from a high-pressure polymorph of lead dioxide. Rietveld refinement of powder x-ray diffraction (XRD), presented in Fig.~\ref{fgr:xray}(b) confirms that the sample is free from impurities. The refined orthorhombic lattice parameters are  $a = 4.816(60)$~\AA, $b = 5.6214(6)$~\AA, and  $c = 5.1199(3)$~\AA, atomic positions and thermal parameters are shown in Table~\ref{tab:structure}. Elemental homogeneity on the micrometer length scale was also confirmed via energy dispersive x-ray spectroscopy (EDS), as shown in Fig.~\ref{fgr:xray}(c). The elemental maps show no evidence for clustering or segregation with all dark regions corresponding to shadows due to the morphology of the powder, as can be compared with the scanning electron micrograph. A more speckled appearance is evident for Hf, which is attributed to the higher energy of the Hf L-edge and the rough sample surface.  

The $\alpha$-PbO$_2$ structure found for (Ti, Zr, Hf, Sn)O$_2$, which is shown in Fig.~\ref{fgr:xray}(d), is an orthorhombic relative of the tetragonal rutile structure. In this structure, all metal cations share a single crystallographic site and sit within a highly distorted octahedral oxygen environment. Whereas in rutile, the local environment is an uniaxially elongated or compressed octahedra with two inequivalent metal-oxygen bond lengths, in the $\alpha$-PbO$_2$ structure, there are three pairs of inequivalent metal-oxygen bonds as shown in Fig.~\ref{fgr:xray}(e). The refined metal-oxygen bond distances in (Ti, Zr, Hf, Sn)O$_2$ vary by more than 10\%, in contrast to the 2\% that is typical for rutile structures. Like rutile, the $\alpha$-PbO$_2$ structure is a network of corner and edge-sharing octahedra. Each oxygen anion is three-fold coordinate and connects a pair of edge-sharing octahedra to the corner of an adjoining octahedra, as shown in Fig.~\ref{fgr:xray}(f). In the rutile structure on the other hand, the edge-sharing octahedra are linked along the $c$ direction.

\begin{table}[tbp]
\caption{Wyckoff sites (Wyck.), lattice coordinates, isotropic thermal parameters ($B_{iso}$), and occupancies (Occ.) for (Ti, Zr, Hf, Sn)O$_2$ in the $\alpha$-PbO$_2$ structure determined by Rietveld refinement. The $R_{wp}$ value is 10.464 and the goodness of fit is 2.27.} 
\noindent
\begin{tabular*}{\columnwidth}{@{\extracolsep{\stretch{1}}}*{7}{c}@{}}
\toprule
\multicolumn{1}{c}{} & Wyck. & $x$       & $y$       & $z$       & $B_{iso}$ & Occ. \\
\hline
Ti                   & 4c    & 0       & 0.316(8) & 0.25    & 1.3(3)      & 0.25 \\
Zr                   & 4c    & 0       & 0.316(8) & 0.25    & 1.3(3)     & 0.25 \\
Hf                   & 4c    & 0       & 0.316(8) & 0.25    & 1.3(3)      & 0.25 \\
Sn                   & 4c    & 0       & 0.316(8) & 0.25    & 1.3(3)      & 0.25 \\
O                    & 8d    & 0.269(9) & 0.108(6) & 0.075(9) & 1.4(1)      & 1  \\ 
\toprule
\end{tabular*}
\label{tab:structure}
\end{table}

The term entropy stabilization is used when there is reason to believe that the thermodynamic stability of a compound over relevant competing phases arises from the entropic contribution to the free energy rather than the enthalpic contribution~\cite{doi:10.1021/jacs.2c11608}. A unique characteristic of (Ti, Zr, Hf, Sn)O$_2$, in comparison to other high entropy oxides, is that it forms into a crystal structure that is not the thermodynamic ground state for any of its constituent elementary oxides. To contrast, in the prototypical rock salt high entropy oxide, three of the five constituents oxides themselves take rock salt as their ground state structure. We can therefore tentatively assign the $\alpha$-PbO$_2$ phase as being entropy stabilized, as the enthalpy preferred states for the constituent oxides are rutile or baddeleyite. Indeed, there is experimental proof that the much simpler oxide with just two components (Ti, Zr)O$_2$ in the $\alpha$-PbO$_2$ structure is already entropy stabilized \cite{Hom2001}. There is also evidence of entropy stabilization from the DFT calculations in this work, as the $\alpha$-PbO$_2$ phase has a positive enthalpy of a magnitude where configurational entropy for four disordered cations overcome the enthalpy barrier at typical synthesis temperatures. Experimental evidence of reversibility is also sometimes used as criteria for entropy stabilization. Our experiments do not show reversibility, which can be attributed to slow diffusion kinetics for the transformation back to constituent oxides at the relevant transformation temperature.

\subsection*{Why the $\alpha$-PbO$_2$ structure?}

We are then left with the question of why the $\alpha$-PbO$_2$ structure is selected as the ground state for (Ti, Zr, Hf, Sn)O$_2$. This structure is unique among high entropy materials for its orthorhombic space group symmetry, where in contrast, the vast majority of high entropy materials with a single cation site crystallize in a cubic high-symmetry crystal structure. We will first attempt to develop an intuitive picture based on qualitative arguments, then follow with detailed DFT calculations in the subsequent sections that validate our picture.

The $\alpha$-PbO$_2$ structure is related to the fluorite and baddeleyite structures through modulations of the oxygen positions. The highest symmetry structure fluorite, followed by the tetragonal zirconia structure, then $\alpha$-PbO$_2$, and finally baddeleyite, are all related by phase transitions consisting of collective displacements of the oxygen sublattice \cite{Troitzsch2005}. Even the continuous transformation to the rutile phase can be imagined \cite{Hyde1972}, involving some bonds breaking and forming. Pauling's first rule derives a minimum ratio for the cation and anion radii in various geometric arrangements. This rule implies that a cation that is too small will lead to an instability and a lower coordination number while a cation that is too large might still be stable in a smaller than ideal anion cage. This principle would imply that the six-coordinated rutile or $\alpha$-PbO$_2$ structures should be favored as soon as any cation preferring a six-coordinate environment is included in the mixture. Given that our four-component compound includes both Sn and Ti, it is thus expected that rutile or $\alpha$-PbO$_2$ should be favored.

\begin{figure*}[htbp]
  \centering
      \includegraphics[width=\textwidth]{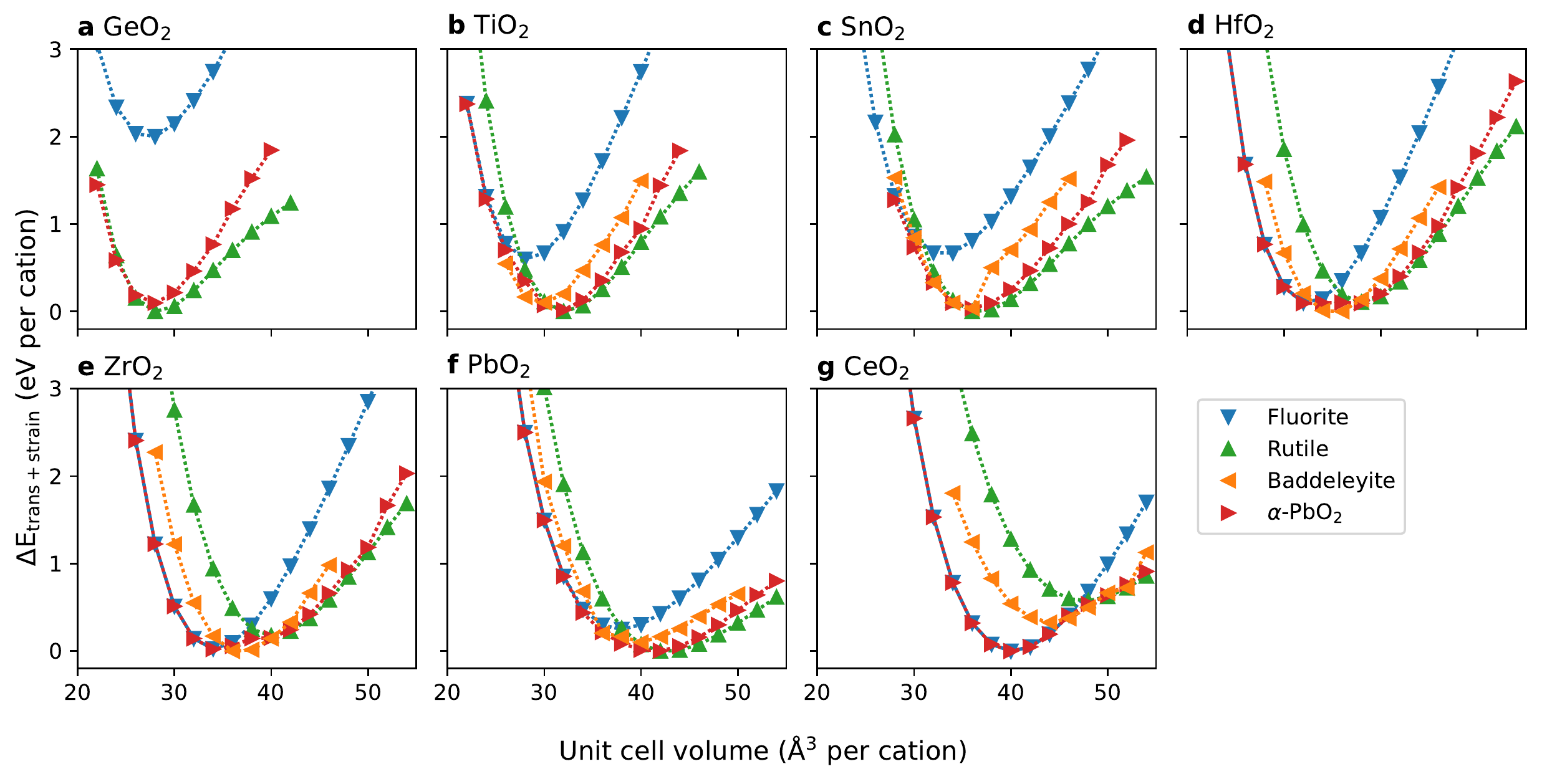}
  \caption{\textbf{Candidate binary oxides under strain.} Energy difference as a function of fixed unit cell volume for the seven $d^0$ and $d^{10}$ tetravalent binary oxides (a) GeO$_2$, (b) TiO$_2$, (c) SnO$_2$, (d) HfO$_2$, (e) ZrO$_2$, (f) PbO$_2$, and (g) CeO$_2$ in the four candidate crystal structures (fluorite, rutile, baddeleyite, and $\alpha$-PbO$_2$). The energy difference $\Delta E_{\text{trans + strain}}$ represents the transformation energy and the strain energy relative to the ground state structure and volume. The cation size is the determining factor both for which crystal structure is most stable (six-, seven-, or eight-fold coordinate) and for which crystal volume the crystal is the most stable. The $\alpha$-PbO$_2$ structure is consistently among the lowest energy structure for all cations and has a relatively flat energy distribution at its minimum, suggesting it can accommodate a larger distribution in cation sizes.} 
  \label{fgr:StrainedBinaries}
\end{figure*}

As discussed previously, there are many similarities between the $\alpha$-PbO$_2$ and rutile structures. In particular, both structures are made up of corner and edge-sharing octahedra and can be viewed as tetragonal close packed oxygen lattices where half of the octahedral voids are filled by cations. While the edge-sharing octahedra in the rutile structure form a chain in the [001] direction, the edge-sharing octahedra in the $\alpha$-PbO$_2$ structure are skewed and form a zig-zag pattern. This lends an intuitive possible explanation for why the $\alpha$-PbO$_2$ structure is favored over the rutile, as it is easier to incorporate local distortions in a zig-zag chain of octahedra as compared to a more rigid straight chain. Additionally, it explains why the $\alpha$-PbO$_2$ structure is a high-pressure phase for the oxides with a rutile ground state. This conceptual understanding is validated through the ensuing DFT calculations.

\subsection*{Phase stability of elemental oxides}

Motivated by the observation that many of the tetravalent oxides under consideration here undergo a pressure-induced transition to an $\alpha$-PbO$_2$ structure, we begin by considering stability of each elemental oxide in relation to its unit cell volume. This is a rational starting point when considering the phase stability of a high entropy phase, as the variation in cation size across the sample can be imagined as exerting a local pressure that expands or contracts the local environment.

As a first order approximation, the enthalpy of an HEO can be determined by the transformation enthalpy from each cations ground state crystal structure to the target crystal structure with the additional local strain or chemical pressure imposed on each cation from a rigid oxygen sublattice of fixed equilibrium volume for all the constituents. Hence, the total energy of the seven oxide binaries in the four relevant crystal structures are calculated for a series of fixed unit cell volumes, as shown in Figure \ref{fgr:StrainedBinaries}, where ionic coordinates and cell shapes are allowed to relax to their minimum energy values. Some data points are omitted for the smallest or largest unit cell volumes when the volumes are so far from equilibrium that the calculations become unstable, especially for the lower symmetry baddeleyite structure which has the most positional degrees of freedom. 

From these quick and simple calculations, some tendencies are already manifesting. The equilibrium volume of the eight-fold coordinated fluorite structure is always smaller than the volumes of the six-fold coordinated rutile and $\alpha$-PbO$_2$ structures, while the seven-fold coordinated baddeleyite structure is usually intermediate. In each case, the lowest energy structure is correctly identified to match the experimentally observed ground state. The lowest energy is found for a unit cell volume that correlates directly to the ionic size of the cation. The largest cation Ce has a preference for the fluorite structure and strongly disfavors the rutile structure, while the smallest cations Ge, Ti, and Sn strongly disfavor the fluorite structure and instead prefer the rutile structure. The $\alpha$-PbO$_2$ and baddeleyite structures have lower curvatures at the bottom of the energy wells than the other structures, indicating that they are able to accommodate cations with a larger distribution of cation radii. In all cases, the $\alpha$-PbO$_2$ structure straddles the most stable regions of both the rutile and the fluorite structure. 

Since the canonical $\alpha$-PbO$_2$ structure is six-fold coordinated, it is expected to be most stable at volumes similar to the six-fold coordinated rutile structure. In practice the volumes are just a bit smaller, reflected in the fact that the $\alpha$-PbO$_2$ structure is typically a high-pressure phase. However, a continuous transformation to a fluorite structure is possible from the $\alpha$-PbO$_2$ structure. Visual inspection of the unit cells where the $\alpha$-PbO$_2$ structure overlaps with the fluorite structure, such as for for ZrO$_2$ at smaller unit cell volumes as shown in Fig.~\ref{fgr:StrainedBinaries}(e), shows that the polyhedra visually look eight-fold coordinated and that the unit cell has a more cubic shape closely resembling the fluorite structure. When the energy barrier between these structures is nonexistent, the $\alpha$-PbO$_2$ structure is not locally stable and DFT relaxations find the higher-symmetry structure. The same scenario occurs for smaller volumes of the baddeleyite structure for SnO$_2$ (Fig.~\ref{fgr:StrainedBinaries}(c)), where the low-symmetry baddeleyite structure is not locally stable and relaxes into an $\alpha$-PbO$_2$ structure. The calculations of Ge in the baddeleyite structure are excluded, because the calculations have noisy energies and a tendency to pull apart and form layered structures. 

From these observations we can already outline some preliminary qualitative expectations: the largest cation Ce will probably not be able to co-exist on the same crystal lattice as the smallest cations Ge and Ti, due to the vast difference in their stable unit cell volume , which is around 40~\AA$^3$/cation in the former case and 30~\AA$^3$/cation in the latter two. In these three cases, the energy landscapes overall are also observed to have the highest curvatures, indicating the lowest tolerance to occupy an environment of differing size. Meanwhile  the intermediate sized cations Sn, Hf, and Zr, are expected to more easily form a solid solution as they all have equilibrium unit cell volumes close to 35~\AA$^3$/cation. The most probable crystal structures for a possible mixture of various subsets of these cations are baddeleyite or $\alpha$-PbO$_2$ due to the lower curvatures around the minimum of their energy wells. 

\begin{figure}
  \centering
  \includegraphics[width=\columnwidth]{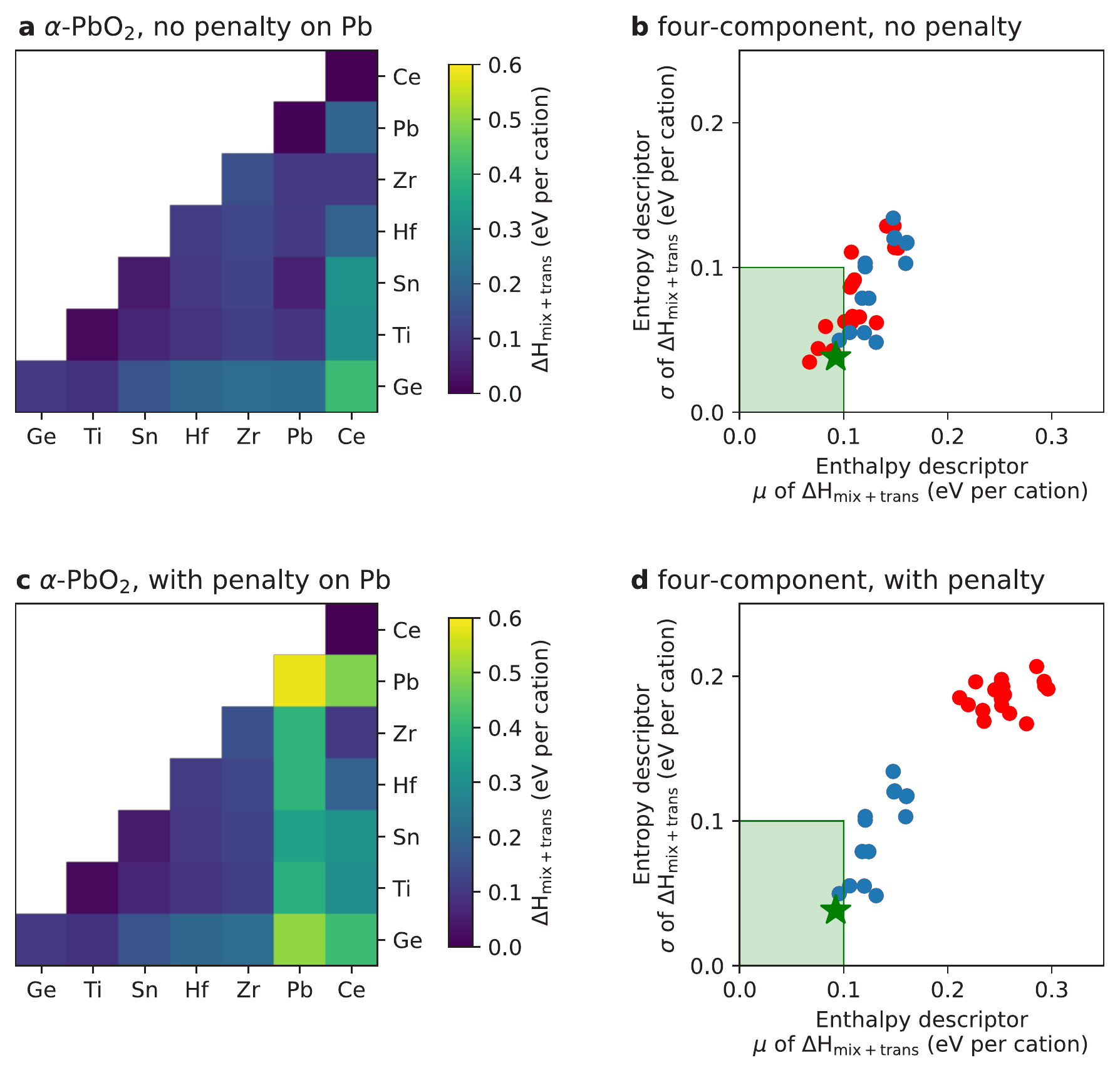}
  \caption{\textbf{Pairwise enthalpies and descriptors for four-component oxides with and without the penalty on Pb.} (a) Heat map showing the $\Delta H_{\text{mix+trans}}[AA', P]$ (enthalpy of mixing and transformation) interaction parameters for all pairs of cations in the $\alpha$-PbO$_2$ structure, representing the favorability of any two cations mixing into a solid solution as compared to their respective ground state energies. (b) A scatter plot of the enthalpy (mean $\mu$) and entropy (standard deviation $\sigma$) descriptors  calculated for the 35 possible four-component oxides. (Ti, Zr, Hf, Sn)O$_2$ is shown as a green star, well within the green region which marks empirical synthesizability. The red points mark Pb-containing compounds, while the remaining blue points represent Ge and/or Ce containing compounds. (c) The same as in (a) but including a penalty of 0.58 eV/Pb atom. (d) Scatter plot as in (b), also including the penalty on Pb. Note that all the Pb-containing compounds are shifted away from stability leaving the green star alone standing out in the green empirical region of stability.}
  \label{fgr:fourcomponent}
\end{figure}

\subsection*{Understanding the phase stability of four-component tetravalent oxides}
In practice, the previous view of the enthalpy is too simplified since cation-cation interactions are not included and some reorganization of the oxygen sublattice is required to accommodate the cations. To explore these effects, special quasi-random (SQS) unit cells of all the 21 possible cation pairs in the four relevant crystal structures are relaxed in a DFT calculation and the energy of these cells are recorded. From these calculations we can determine an interaction energy, $\Delta H_{\text{mix+trans}}[AA', P]$, which represents the mixing enthalpy between pairs of cations $AA'$ in addition to the enthalpy of transformation from the ground state to the crystal structure $S$ (details can be found in the methods section). The lower the interaction energy, the closer the energy of the mixture is to the ground state of the independent binary oxides. The heat map showing the interaction parameters for all the cation pairs in the $\alpha$-PbO$_2$ structure is shown in Figure \ref{fgr:fourcomponent}(a), which is the experimentally observed structure for (Ti, Zr, Hf, Sn)O$_2$. The axes are the $A$ and $A'$ cations sorted by their Shannon radii. It shows that most cation pairs are accommodated well in this crystal structure having interaction energies close to zero, except the smallest ion (Ge) and the largest ion (Ce). The more dissimilar the cation radii are, the higher their mixing enthalpy is, again as expected and visible from the lower energy toward the bottom left-top right diagonal as opposed to the higher energies in the bottom right corner. Although some trends in these heat maps are as expected, it is important to emphasize that the ternary interactions are more than just a linear combination of the binary oxides.

\begin{figure*}
  \centering
  \includegraphics[width=\textwidth]{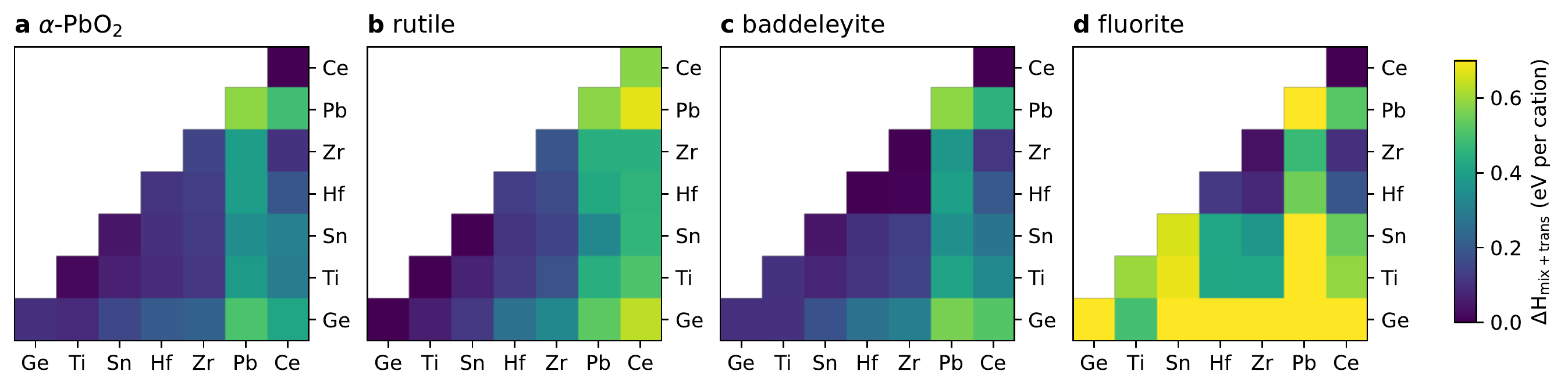}
  \caption{\textbf{Pairwise mixing and transformation enthalpies.} Heat maps of the $\Delta H_{\text{mix+trans}}[AA', P]$ interaction parameters for the 21 possible ternary oxides in the four different crystal structures: (a) $\alpha$-PbO$_2$, (b) rutile, (c) baddeleyite, and (d) fluorite, representing the favorability of any two cations mixing into a solid solution as compared to their respective ground state energies. These maps include the energetic penalty on Pb for its reduction potential at high temperatures. Most of the Ge and Pb containing tiles in panel (d) are saturated, they are energetically unfavorable with high transformation energies from the ground state to the fluorite structure with coefficients as high as 2.32 eV. The color axis is chosen to allow the smaller energy scales closer to 0 to be visibly resolved. The fluorite structure is strongly disfavored for all combinations of cations except for Ce with itself. The other three structures appear more similar in their favorability with the exception of Pb and Ce. }
  \label{fgr:heatmaps}
\end{figure*}

Following Pitike \textit{et al.} \cite{Pitike2020}, the average, $\mu$, of all relevant $\Delta H_{\text{mix+trans}}[AA', P]$ for any possible combination of four cations in the $\alpha$-PbO$_2$ structure is used as an enthalpy descriptor, while the standard deviation of the same quantity, $\sigma$, is used as an entropy descriptor. Calculating these values for all $\binom{4}{7} = 35$ four-component compositions results in a scatter plot as shown in Figure \ref{fgr:fourcomponent}(b). The four-component oxide already synthesized, (Ti, Zr, Hf, Sn)O$_2$, is marked as a green star. The green square enclosing the area of 0.1 eV/cation in both the enthalpy and entropy descriptor axes shows an empirical region where synthesis may be possible based on the observations of Pitike \textit{et al.}~\cite{Pitike2020}. Closely surrounding (Ti, Zr, Hf, Sn)O$_2$ are four combinations where one of the four cations (Ti, Zr, Hf, Sn) are replaced by Pb. All Pb-containing compositions are marked with red. The compounds that contain both Ge and Ce are disfavored, while the compounds that contain either are at the border of the empirical synthesis boundary.

A major thermodynamic feature that was not included in the strained binary calculation or yet in the pairwise calculations is the fact that the Pb cation has a preference for the 2+ oxidation state over the 4+, at typical synthesis temperatures. Experimental thermodynamic data can be used to calculate an energetic penalty for the preferred oxidation state of Pb, which is found to be 0.58 eV/Pb at 1159 K. The experimental foundation, calculations, and justifications for this penalty are outlines in detail in the Methods section. By including this penalty, all the Pb-containing tiles in the heat map in \ref{fgr:fourcomponent}(a) are shifted significantly upwards in energy as shown in \ref{fgr:fourcomponent}(c). As a result, the Pb-containing four-component compounds marked with red in Figure \ref{fgr:fourcomponent}(b) shift to higher entropy and enthalpy values, and the star indicating our four-component synthesis stands alone as the most likely stable compound in Figure \ref{fgr:fourcomponent}(d), with an enthalpy descriptor of 0.092 eV/cation and an entropy descriptor of 0.038 eV/cation. The small and positive enthalpy value is an indicator of entropy stabilization, a value which is balanced with the configurational entropy associated with disordering of four cations at a temperature of 773 K, a threshold we substantially surpass in our synthesis (maximum temperature 1773~K). This can be an explanation why the transformation back to binary oxides is kinetically hindered. For completeness, the four-component enthalpy and entropy descriptors for all the four crystal structures are shown both with and without the energy penalty in Supplementary Figures 1 and 2.

\subsection*{Predicting the phase stability of five-component tetravalent oxides}
Motivated by our success in understanding the phase stability of the four-component entropy stabilized oxide, we next attempt to find a stable five-component material. All pairwise parameters are shown as heat maps in Figure \ref{fgr:heatmaps} for all the four candidate crystal structures. As before, these maps include the penalty on Pb due to its tendency to reduce from 4+ to 2+ at high temperatures. The corresponding maps excluding this penalty can be found in Supplementary Figure 3. From these heat maps, it is again clear that the fluorite structure is disfavored for the smaller cations Ti, Sn, and Ge; while the largest cation Ce is disfavored in the remaining structures, as expected from the strained binary calculations.

\begin{figure*}
  \centering
  \includegraphics[width=\textwidth]{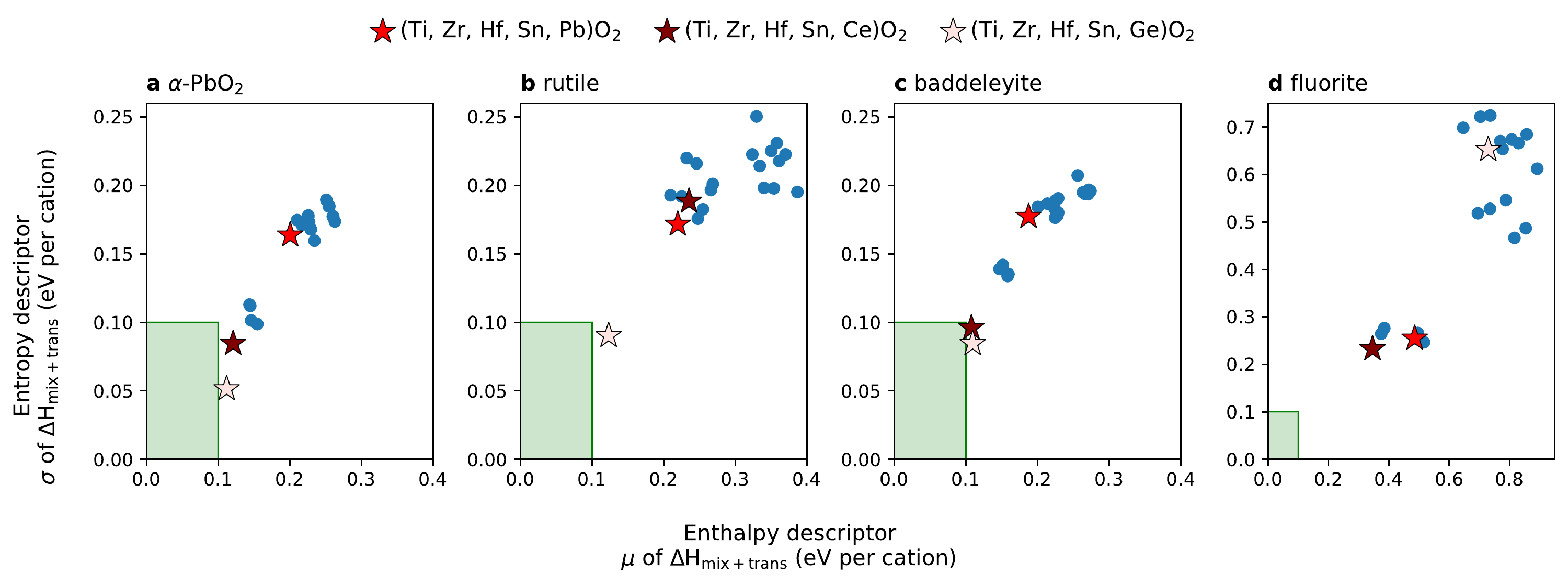}
  \caption{\textbf{Entropy and enthalpy descriptors for five-component oxides.} Entropy and enthalpy descriptors ($\mu$ and $\sigma$ of $\Delta H_{\text{mix+trans}}[AA', P]$ respectively) including the penalty on Pb for all the possible five-component oxides in the four crystal structures: (a) $\alpha$-PbO$_2$, (b) rutile, (c) baddeleyite, and (d) fluorite. The green shaded region indicates empirical synthesizability. Most compositions are far outside of the green region, but some of our attempted syntheses are quite close. The three compositions marked with stars, (Ti, Zr, Hf, Sn, Pb)O$_2$ (red star), (Ti, Zr, Hf, Sn, Ce)O$_2$ (maroon star), and (Ti, Zr, Hf, Sn Ge)O$_2$ (pink star) are identical to the successfully synthesized four-component phase with the addition of Pb, Ce, and Ge, respectively. We find that none of these compositions form as single phase materials under solid state synthesis conditions, supporting the empirical range of stability for the enthalpy and entropy descriptors.}
  \label{fgr:fivecomponent}
\end{figure*}

In order to quantify which five-cation oxides are most likely to be stable, the descriptors are calculated from the pairwise mixing enthalpies for $\binom{5}{7} = 21$ compounds in the four crystal structures. These descriptors are shown as scatter plots in Figure \ref{fgr:fivecomponent}. For all possible compositions, the fluorite structure is far from stability, hence the different scale for that panel while the size of the green area is the same. No five-component oxide is found well within the green area for any structure type, although some compounds are just outside the border of stability. The ones at the border are the ones where we add either the smallest cation Ge or the largest cation Ce to the existing four-component mixture, (Ti, Zr, Hf, Sn, Ge)O$_2$ and (Ti, Zr, Hf, Sn, Ce)O$_2$, marked by pink and maroon stars, respectively. The composition that includes Ge is also close to stability for both the rutile and baddeleyite structures but the entropy and enthalpy descriptors are both miminized in the $\alpha$-PbO$_2$ structure. A similar scenario occurs for the Ce containing composition in the baddeleyite structure. Therefore, if a single phase material can form for these borderline compositions, we expect it will have the $\alpha$-PbO$_2$ structure. 

If the penalty on Pb is not included, however, the compound (Ti, Zr, Hf, Sn, Pb)O$_2$ has the lowest value for both the enthalpy and entropy descriptors, and is expected to crystallize in the $\alpha$-PbO$_2$ structure. The descriptors are shown in Supplementary Figure 4, and the values are similar to those of the four-component compound, indicating that this compound should be synthesizeable if high temperature is avoided. The $\alpha$-PbO$_2$ and the baddeleyite panels are similar except the larger magnitude of the entropy descriptor for baddeleyite, and therefore the $\alpha$-PbO$_2$ phase is favored for this composition. When the penalty on Pb is included, this composition (Ti, Zr, Hf, Sn, Pb)O$_2$ is far from stability, as marked with a red star in Fig.~\ref{fgr:fivecomponent}.

\subsection*{Attempts at synthesizing a five-component tetravalent oxide}

To validate the picture developed through our computational approach, we have attempted the synthesis of several five-component oxides. All of our compositions involve the same constituents as our single phase four-component material (Ti, Zr, Hf, Sn)O$_2$, plus the addition of a fifth element (Ge, Pb, or Ce). None of these are straightforwardly expected to form a single phase material based on the signatures of phase stability in our entropy and enthalpy descriptors approach. Both the Ge- and Ce-containing compositions are just outside the enthalpy boundary of the approximate stability field for the $\alpha$-PbO$_2$, rutile (only for Ge), and baddeleyite structures, while the Pb-containing compositions are far from stability once accounting for the reduction potential for PbO$_2$ to PbO. In our attempts, none of these compositions led to the formation of a single phase high entropy oxide, as we will discuss in detail below. This supports the validity of the empirical boundaries on phase stability of $\mu=0.1$~eV/cation and $\sigma=0.1$~eV/cation.

Our synthesis attempt of the Ge-containing composition (Ti, Zr, Hf, Sn, Ge)O$_2$ followed the same solid state synthesis procedure as the four-component compound. However, the nearly coincident melting (1115~$^\circ$C) and boiling points (1200~$^\circ$C) of GeO$_2$ create a strict upper bound on the reaction temperature. To limit evaporation of GeO$_2$ a reaction temperature of 1050~$^\circ$C was selected. Extended reaction at this temperature resulted in a multi-phase material containing a mixture of a cation ordered zirkon-like HfGeO$_4$ phase (space group $I41/a$, no. 88), a rutile phase, and an $\alpha$-PbO$_2$ phase, with additional annealing time favoring a large fraction of the HfGeO$_4$ phase. This phase is isostructural with a ZrGeO$_4$ phase, and solid solubility is expected. A representative phase matched x-ray diffraction pattern for this failed synthesis is shown in Supplementary Figure 5. This incomplete reaction is unsurprising as high reaction temperatures are needed to overcome kinetic barriers for the refractory oxides. Therefore, we also attempted to gradually increase the synthesis temperature closer to and beyond the boiling point of GeO$_2$. This synthesis route led to the nearly complete evaporation of GeO$_2$, as verified by quantitative elemental analysis with EDS.

Multiple attempts were also made to synthesize the Ce-containing composition (Ti, Zr, Hf, Sn, Ce)O$_2$. We note that this is an identical composition to a previously reported single phase fluorite high entropy oxide~\cite{Chen2018}. Our extensive attempts to replicate this result following their method did not yield a single phase material but instead produced a mixture of fluorite and $\alpha$-PbO$_2$. A representative diffraction pattern for a sample reacted at 1500~$^\circ$C reaction temperature is shown in Supplementary Figure 5. Similar results were obtained for additional annealing time and annealing at higher temperatures (max 1600~$^\circ$C). These findings are challenging to reconcile with those of Chen \textit{et al.}~\cite{Chen2018}, particularly in light of their elemental mapping that shows a complete phase segregation of Ce on a micron lengthscale within a matrix of Ti, Zr, Hf, and Sn, which our work demonstrably shows form as an $\alpha$-PbO$_2$ phase. We are not aware of any other works that have successfully reproduced this result. It is interesting to note that both our intuitive picture and our DFT analysis would suggest that this composition would be highly unstable in the fluorite phase. Both Ti and Sn are significantly too small for the rigid eight-fold coordinate environment of the fluorite structure and their incorporation into a fluorite phase would violate Pauling's first rule. This can be confirmed by the phase stability for the elemental oxides shown in Figure~\ref{fgr:StrainedBinaries}, where both TiO$_2$ and SnO$_2$ have ground state energies that are more than 1~eV/cation larger than the three other structures formed here. One can also consider the $\Delta H_{\text{mix+trans}}[AA', P]$ for various combinations of elements in the fluorite structure, as shown in Fig.~\ref{fgr:heatmaps}(d), where we see that for almost all pairs of cations the fluorite structure is highly unfavorable, except for pure CeO$_2$. If (Ti, Zr, Hf, Sn, Ce)O$_2$ does exist as a single phase material, we would expect it to form in the $\alpha$-PbO$_2$ phase based on the calculated descriptors. Other fluorite high entropy oxides have been reported but these typically involve other rare earth cations with an oxidation state of 3+~\cite{gild2018high,wright2020high} leading to oxygen vacancies which can help stabilize the fluorite structure. Notably, Ca$^{2+}$-substituted varieties of the (Ti, Zr, Hf, Sn, Ce)O$_2$ compound do not show segregation of Ce on a micron scale, again highlighting the importance of oxygen vacancies in addition to the presence of Ce in stabilizing the high entropy fluorite structure, even enabling for the inclusion of Ti$^{4+}$ and Sn$^{4+}$ cations~\cite{Chen2023Casubstituted}.

Given the name of the structure, $\alpha$-PbO$_2$, one would naturally suspect that Pb is the most likely candidate for a fifth element to incorporate. The $\alpha$-PbO$_2$ phase was originally discovered as a deposit formed on the surface of the anode of a lead battery~\cite{zaslavskii1950new}. This metastable structure of PbO$_2$ was later discovered to also form under high pressure conditions at relatively low temperatures~\cite{white1961high}. Our calculations do indeed suggest that (Ti, Zr, Hf, Sn, Pb)O$_2$ is the most stable five-component tetravalent oxide of the elements we surveyed and would be expected to form in the $\alpha$-PbO$_2$ structure with one significant caveat: the stability of the oxidation state of the ion in question. While for all other ions except Pb, the 4+ oxidation state is the most stable at all relevant temperatures, for Pb, under ambient conditions reduction towards the 2+ state occurs starting from 300~$^\circ$C and is complete at 600~$^\circ$C. Therefore, this reduction occurs significantly sooner than the formation of the $\alpha$-PbO2 phase for (Ti, Zr, Hf, Sn)O$_2$, which can observed to start forming around 1150~$^\circ$C (Fig.~\ref{fgr:xray}(a)). Despite this obstacle, we did attempt the synthesis of (Ti, Zr, Hf, Sn, Pb)O$_2$ following two different reaction pathways. In the first, all five reagent oxides were combined while in the second, a precursor of the four-component (Ti, Zr, Hf, Sn)O$_2$ was combined with PbO$_2$. In both cases, no heat treatment was found to successfully produce a single phase five-component phase without significant mass loss or the formation of a secondary cation ordered perovskite phase involving lead in its reduced form, as shown for the x-ray diffraction pattern in the Supplementary Figure 5. 

While none of the five-component tetravalent oxides could be successfully obtained here via solid state synthesis, there are alternative synthesis methods that may prove fruitful, particularly in the case of Ge and Pb. In the case of Ge, the barrier to be overcome is the evaporation of GeO$_2$, which could possibly be suppressed under high-pressure conditions. High-pressure can also speed up reaction kinetics and may promote the complete transformation to the $\alpha$-PbO$_2$ phase at a slightly lower reaction temperature. In the case of Pb, the major barrier to be overcome is the reduction of PbO$_2$ to PbO and the resulting formation of cation ordered phases. Under different synthesis conditions, such as the strongly acidic solution environment of a lead battery, it is possible to achieve Pb in the 4+ oxidation state as the thermodynamically stable option. One way to influence the relative stability of PbO$_2$ and PbO is to perform the synthesis in high oxygen partial pressure conditions. However, by linear extrapolation of the phase diagram in Figure 17 in Risold \textit{et al.} \cite{Suzuki1998} an oxygen partial pressure at the order of 100~GPa might be necessary to avoid reduction at expected solid state synthesis temperatures. There is also the low melting point of the Pb-based oxides to consider, PbO melts 500$^\circ$C below the synthesis temperature for the (Ti, Zr, Hf, Sn)O$_2$. Therefore a solution-based method or mechanochemical synthesis, avoiding the high temperatures required for solid state synthesis, may be the most promosing synthesis routes. Another alternative route to achieve a single-phase five-component material is to adjust the stoichiometry of the composition to off-equimolar ratios, lowering the amount of the elements with the highest transformation energies.

\section*{Conclusions}
In this work we explored the phase stability of high entropy mixtures of tetravalent $d^0$ and $d^{10}$ cations. We replicated the synthesis of the four-component entropy stabilized oxide (Ti, Zr, Hf, Sn)O$_2$, which crystallizes in the $\alpha$-PbO$_2$ structure. A unique characteristic of this entropy stabilized phase is that it forms in a low symmetry orthorhombic phase and that the resulting structure is distinct from the ground state structure of any of its binary oxides. The geometry of the $\alpha$-PbO$_2$ structure and its relationship to the competing phases can be used to justify why this structure is selected for this phase. The effect of strain and pairwise cation interactions are studied using DFT, solidifying the understanding of why (Ti, Zr, Hf, Sn)O$_2$ forms in the $\alpha$-PbO$_2$ structure. When the reduction of Pb$^{4+}$ to Pb$^{2+}$ at synthesis temperatures is accounted for, the calculated enthalpy and entropy descriptors indicate that this is the only stable four- or five-component oxide in this family of oxides. This result is validated by the attempted synthesis of several five-component oxides, which fail to yield a single phase material, in accordance with the calculations. Our work highlights the promise of a combined computational and experimental approach in the quest to discover new HEOs.

\section*{Methods}
\subsection*{Experimental methods} All samples were prepared by solid state synthesis. Binary oxides were mixed in ethanol in an agate mortar, uniaxially pressed to pellets, and heat treated in multiple steps with increasing temperatures and intermediate regrindings to assess the onset of the reaction temperature. Following each heat treatment the samples were quenched in air. The pellets were weighed before and after heat treatments to monitor weight loss. Synthesis of (Ti, Zr, Hf, Sn, Ge)O$_2$, (Ti, Zr, Hf, Sn, Pb)O$_2$, and (Ti, Zr, Hf, Sn, Ce)O$_2$ were also attempted, as outlined in the text. 

Phase purity of the synthesized materials was assessed using powder x-ray diffraction (XRD). Measurements were performed on a Bruker D8 Advance diffractometer using a Cu x-ray source with monochromated $\lambda_{K\alpha_1} = 1.5406$~\AA~in Bragg-Brentano geometry. Rietveld refinements were performed using TOPAS~\cite{coelho2018topas} and refining the background, lattice constants, zero error, atomic positions, simple axial model, TCHZ peak shape, strain broadening, and thermal parameters. Elemental homogeneity of the synthesized materials was evaluated using energy dispersive x-ray spectroscopy (EDS) using a Philips XL30 electron microscope equipped with the Bruker Quantax 200 energy-dispersion X-ray microanalysis system. 

\subsection*{Density functional theory calculations} The Vienna Ab initio Simulation package~\cite{Kresse1993, Kresse1994, Kresse1996Efficiency, Kresse1996Efficient} (VASP) version 5.4.4 with the projector augmented wave~\cite{Blochl1994, Kresse1999} (PAW) method was used for the Density Functional Theory (DFT) calculations. The VASP supplied PBE~\cite{PBE1996} PAW potentials version 5.4 were used, including 12 electrons for Ti, Zr, Hf, and Ce, respectively 14 electrons for Ge, Sn, and Pb in the valence shells. The PBEsol+U~\cite{PBEsol2008} exchange correlation functional was used, with Hubbard U values of 4.35, 3.35, and 2.7 for the d-electrons on Ti, Zr, and Hf respectively; 0 eV for Ge, Sn, and Pb; and 4.0 eV for the f-electrons on Ce, calculated by the linear response approach~\cite{Cococcioni2005} and confirmed to finding the correct ground state for all elements. This correction is necessary to get the correct ground state for TiO$_2$ which prefers the $\alpha$-PbO$_2$ structure over the experimentally observed rutile without this correction \cite{Curnan2015}. An energy cutoff of 700 eV was used, together with an electronic convergence threshold of 10$^{-8}$ eV and a threshold of maximum force of any one ion of 10$^{-4}$ eV/{\AA}. Gaussian smearing with a width of 0.01 eV and $\Gamma$-centered k-meshes with a spacing of 0.4 {\AA}$^{-1}$ were used. The atomic simulation environment was used to manage the large amount of calculations~\cite{ase-paper}.

\subsection*{Mixing and transformation enthalpies} Mixing and transformation enthalpies were calculated following Pitike et al. \cite{Pitike2020}, 
\begin{multline}
    \Delta H_{\text{mix+trans}}[AA', P] = E_{\text{DFT}}[AA',P] \\-  0.5E_{\text{DFT}} [A,G]-0.5E_{\text{DFT}} [A',G],
    \label{equation}
\end{multline}

where $\Delta H_{\text{mix+trans}}[AA', P]$ is the mixing enthalpy of a pair of cations cations A and A' in any phase P and their transformation energy from the ground state structure to said phase. The energies from three DFT calculations is needed, $E_{\text{DFT}}[AA',P]$ represents the target phase with A an A' mixed (the approach used to calculated these energies are described in the next paragraph), while $E_{\text{DFT}} [A,G]$ and $E_{\text{DFT}} [A',G]$ represent the ground state of the constituent $A$O$_2$ and $A'$O$_2$ oxides. The average value of $\Delta H_{\text{mix+trans}}[AA', P]$ of all cation pairs in a high entropy compound, $\mu$, is used as a descriptor of the enthalpy, while the standard deviation, $\sigma$, of the same quantity is used as the entropy descriptor. The pairwise calculations were run allowing symmetry allowed atomic positions and unit cell volume to relax.

\subsection*{Special quasi-random structures} Because of the more complex topology of the present crystal structures in this work compared to the ones investigated by Pitike \textit{et al.}, special quasi-random (SQS) cells were used to capture all nearest cation neighbour (that is all the cations sharing at least one oxygen bond) orderings equally in the mixed cation calculations ($E_{\text{DFT}}[AA',P]$). This approach was used based on the realization that the unit cell used by  Pitike \textit{et al.} is the smallest possible cell that matches all the nearest cation neighbour correlations. SQS cells were generated using the gensqs and mcsqs codes which are part of the Alloy Theoretic Automated Toolkit (ATAT)~\cite{VanDeWalle2013}. The smallest possible cell to achieve a perfect match to the objective function as defined in ATAT for the nearest cation neighbours  is 6 cations for the fluorite structure and 24 atoms for the remaining structures. There is only one SQS cell that satisfies these conditions for the fluorite structure, and 8 to 20 SQS cells for the remaining structures. All these 43 enthalpies were calculated for the (Ti,Zr)O$_2$ case, and the standard deviation was 6 to 8 meV/atom depending on the crystal structure. Just one version of the SQS cells was used for the remaining calculations.

\subsection*{Reduction penalty on Pb} An extra energy penalty can be added if the ground state does not have the same oxidation state as the target phase, which could be relevant for Pb in this work. Thermodynamic investigation of the Pb-O system reveal that rutile PbO$_2$ is the thermodynamically favored compound in ambient conditions\cite{Suzuki1998}, however, reduction of PbO$_2$ starts at 502 K and proceeds through a few intermediate phases to massicot PbO at 807 K. Solid state reactions are performed at high temperatures where PbO is favored. In the case of Pb, equation \ref{equation} must be adjusted because the reaction it is based on - \mbox{1/2AO$_2$ + 1/2PbO$_2$ $\rightarrow$ (A,Pb)$_2$} - must be replaced with \mbox{1/2AO$_2$ + 1/2PbO + 1/4O$_2$ $\rightarrow$ (A,Pb)$_2$} to represent the correct ground state. The difference between using the 1/2PbO$_2$ and 1/2PbO + 1/4O$_2$ as reference states states can be found from experimental thermodynamic data, and this difference can simply be added every time a Pb atom is included in the calculation. Using the Gibbs free energies for PbO and PbO$_2$ from \cite{Suzuki1998} and for O$_2$ from \cite{Dinsdale1991}, the $\Delta$G of the reaction \mbox{PbO$_2$ $\rightarrow$ PbO + 1/2O$_2$} was found to be -55.7 kJ/mol at 1159 K, which is the melting point of PbO. This corresponds to an energy penalty of 0.58 eV/Pb$^{4+}$ atom, which can be added as a correction to account for the reduction potential of Pb. The remaining elements in this study are stable in the 4+ oxidation state at the relevant temperatures. The reason why experimental data is used instead of DFT calculated energies is that the tendency of DFT to overbind the oxygen molecule and the corrections that follow which tend to make these calculations unreliable. If we were to use the energies (and corrections) from the Materials Project, the penalty would be 0.51 V/Pb which is close to the experimental value we find. However, this value is supposed to be representative of a 0 K scenario and shows that DFT finds he wrong ground state for the Pb-O system.

\section*{Data availability}
The data that support the findings of this study are available from the corresponding author upon reasonable request.

\begin{acknowledgments}
The authors thank Jacob Kabel at the Electron Microbeam \& X-Ray Diffraction Facility at the Department of Earth, Ocean and Atmospheric Sciences at the University of British Columbia for assistance collecting the energy dispersive x-ray spectroscopy data. The authors also thank the members of the Quantum Matter Institute's ``Atomistic approach to emergent properties of disordered materials'' Grand Challenge for insightful conversations on high entropy oxides. This work was supported by the Natural Sciences and Engineering Research Council of Canada and the CIFAR Azrieli Global Scholars program. This research was undertaken thanks in part to funding from the Canada First Research Excellence Fund, Quantum Materials and Future Technologies Program. 

\end{acknowledgments}

\section*{Author contributions}

This study was initiated and supervised by A.M.H. Solid state synthesis and x-ray diffraction was performed by G.H.J.J. and S.M., and analyzed by S.S.A. EDS measurements were performed by S.S.A. and M.O. DFT calculations were performed by S.S.A. under the guidance of J.R. The manuscript was written by S.S.A. and A.M.H. with input from all authors.

\section*{Supplementary information}
Supplementary information contains five figures including heat maps for the four crystal structures excluding the penalty on Pb, enthalpy and entropy descriptors for the five-component compounds excluding the penalty on Pb, enthalpy and entropy descriptors for the four-component compounds including and excluding the penalty on Pb, and x-ray diffractograms with phase identification of the failed five-component syntheses attempts.

\bibliography{bibliography}

\end{document}